# Photoluminescence of lead-related optical centers in single-crystal diamond

S. Ditalia Tchernij[1,2], T. Lühmann[3], J. Forneris[2,4*], T. Herzig[3], J. Küpper[3], A. Damin[5], S. Santonocito[1], P. Traina[6], E. Moreva[6], F. Celegato[6], S. Pezzagna[3], I.P. Degiovanni[6], M. Jakšić[4], M. Genovese[6,2], J. Meijer[3], P. Olivero[1,2]

[1]*Physics Department and "NIS" Inter-departmental Centre - University of Torino, Italy*
[2]*Istituto Nazionale di Fisica Nucleare (INFN), Sez. Torino, Italy*
[3] *Department of Nuclear Solid State Physics, University Leipzig, Germany*
[4] *Ruđer Bošković Institute, Zagreb, Croatia*
5 *Chemistry Department and"NIS" Inter-departmental Centre - University of Torino, Italy*
[6]*Istituto Nazionale di Ricerca Metrologica (INRiM), Italy*

\* corresponding author. Email: forneris@to.infn.it

## Abstract
We report on the creation and characterization of Pb-related color centers in diamond upon ion implantation and subsequent thermal annealing. Their optical emission in photoluminescence (PL) regime consists of an articulated spectrum with intense emission peaks at 552.1 nm and 556.8 nm, accompanied by a set of additional lines in the 535700 nm range. The attribution of the PL emission to stable Pb-based defects is corroborated by the correlation of its intensity with the implantation fluence of Pb ions, while none of the reported features is observed in reference samples implanted with C ions. Furthermore, PL measurements performed as a function of sample temperature (143-300 K range) and under different excitation wavelengths (532 nm, 514 nm, 405 nm) suggest that the complex spectral features observed in Pb-implanted diamond might be related to a variety of different defects and/or charge states.

This work follows from previous reports on optically active centers in diamond based on group IV impurities, such as Si, Ge and Pb. In perspective, a comprehensive study of this set of defect complexes could bring significant insight on the common features involved in their formation and opto-physical properties, thus offering a solid basis for the development of a new generation of quantum-optical devices.

## Main text

Diamond is a promising platform for the development of solid state quantum devices with applications in quantum information processing and sensing [1-5]. In recent years, the search for optically active defects with appealing properties has led to the discovery of several classes of color centers [6-11] alternative to the widely investigated negatively charged nitrogen-vacancy complex (NV⁻ center) [12]. This is motivated by the fact that the NV⁻ defect, although extremely promising for its unique spin features [13], is limited in several applications by suboptimal opto-physical properties such as a spectrally broad emission with intense phonon sidebands, the presence of charge state blinking, and a relatively low emission rate [14-16]. Particularly, the silicon-vacancy center (SiV) [17] has attracted significant attention due to a near transform-limited emission [18,19], good photon indistinguishability [19], the capability of coherently control its spin properties [21-24] and the emergence of reliable techniques for its deterministic fabrication [25-28].

The recent exploration of additional emitters related to group-IV impurities, such as the germanium-related (GeV) [29-31] and tin-related (SnV) color centers [32,33], characterized by similar defect structure and opto-physical properties to those of the SiV center (photo-stability, narrow zero-phonon line - ZPL, relatively small phonon coupling, high emission rate) brings further interest in this class of optically active defects. In particular, these recent discoveries open the question on whether the whole set of group-IV-based complexes (SiV, GeV, SnV, and now PbV) result in stable optically-active emitters, and secondly whether the properties of these color centers exhibit any common features.

In this work, we present the evidence of photoluminescence (PL) emission from color centers created upon the implantation of Pb ions in diamond followed by thermal annealing. To the best of the authors' knowledge, in a recent work [34] promisingly similar results were obtained from low-fluence Pb implantations in diamond, although with several differences which could be ascribed to different experimental conditions (excitation wavelengths, spectral filtering, measurement temperature).

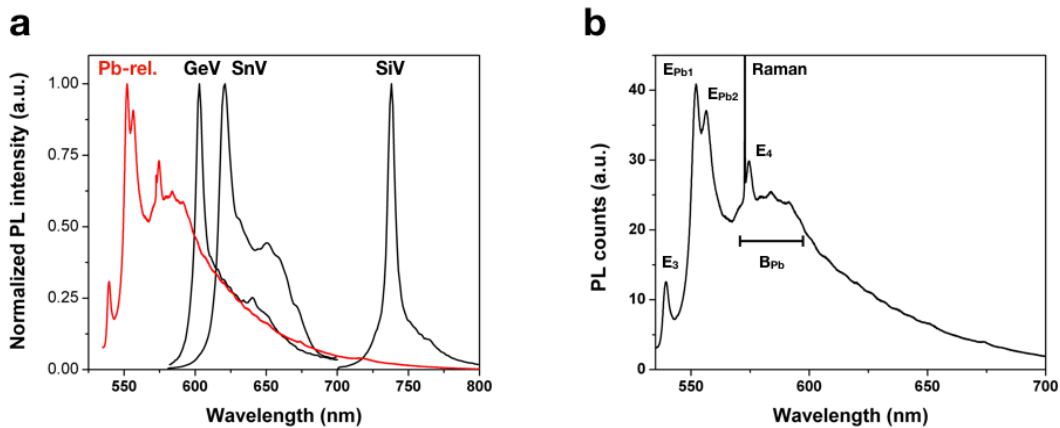

**Figure 1: a)** PL emission spectra from group-IV-related impurities upon ion implantation in the 1-10×10$^{12}$ cm$^{-2}$ fluence range: SiV (ZPL at 738 nm), GeV (602 nm), SnV (620 nm) and the newly reported Pb-related (main emission line at ~552 nm, red curve) centers. **b)** Typical PL spectrum of a detector grade diamond substrate (Sample #1) implanted with 30 keV PbO$_2^-$ ions at a 2×10$^{13}$ cm$^{-2}$ fluence.

Samples processing. The defects fabrication was performed by implanting 35 keV PbO$_2^-$ and 20 keV Pb$^-$ ions at different fluences in two CVD-grown single-crystal IIa diamond substrates supplied by ElementSix. The substrates were characterized respectively by "electronic" (substitutional N and B concentration: [N$_S$] < 5 ppb, [B$_S$] < 5 ppb, Sample #1 in the following) and "optical" ([N$_S$] < 1 ppm, [B$_S$] < 0.05 ppm, Sample #2) grades. The utilization of a PbO$_2^-$ beam was motivated by the low electron affinity of Pb, resulting in relatively small currents achievable for the implantation of elemental ions [35]. The chosen molecular beam thus ensured the achievement of relatively higher implantation fluences for the ensemble characterization of Pb-related defects. Additional implantations of group-IV related elements were performed to provide a suitable comparison in the present study: 50 keV Si$^-$ (5×10$^{12}$ cm$^{-2}$ ion fluence, Sample #1), 40 keV Ge$^-$ (1×10$^{13}$ cm$^{-2}$, Sample #2) and 60 keV Sn$^-$ (1×10$^{12}$ cm$^{-2}$, Sample #1). The implantation was followed by a thermal annealing at 1200 °C in vacuum (2 hours) and by an oxygen plasma treatment, which ensured the removal of background fluorescence associated with graphitic and organic contaminations of the samples surface.

Room-temperature PL emission. A typical room-temperature PL spectrum acquired in confocal microscopy from an ensemble of centers in Sample #1 (2×10$^{13}$ cm$^{-2}$ implantation fluence) under 532 nm laser excitation is reported in **Fig. 1** (red line), together with those of SiV (ZPL at 738 nm), GeV (602 nm), and SnV (620 nm) centers (black lines). The most apparent spectral features associated with Pb implantation (**Fig. 1b**) consist of an intense doublet at 552.1 nm (E$_{Pb1}$ peak in the following) and 556.8 nm (E$_{Pb2}$), followed by a broader emission band (B$_{Pb}$) roughly comprised between 565 nm and 600 nm. These spectral structures show a striking resemblance with that observed for the SiV, GeV and SnV centers, where the emission is mainly concentrated at the ZPL, with a less pronounced phonon sideband. Indeed, a common trend in the reported spectra is represented by the fact that the intensity of the phonon sidebands increases with for heavier impurities (**Fig. 1a**). Furthermore, the heaviest known group-IV-related center reported before the present work, i.e. the SnV defect, also displays a ZPLs doublet fine structure, indicating that the center could have a split ground (and/or excited) state. Apart from this, it is worth remarking that the ZPL wavelength does not follow a systematic red/blue-shift when considering progressively heavier impurities. In addition to the afore-mentioned spectral features, less intense sharp lines are observed at 539.4 nm and 574.5 nm (E$_3$ and E$_4$ peaks in the following).

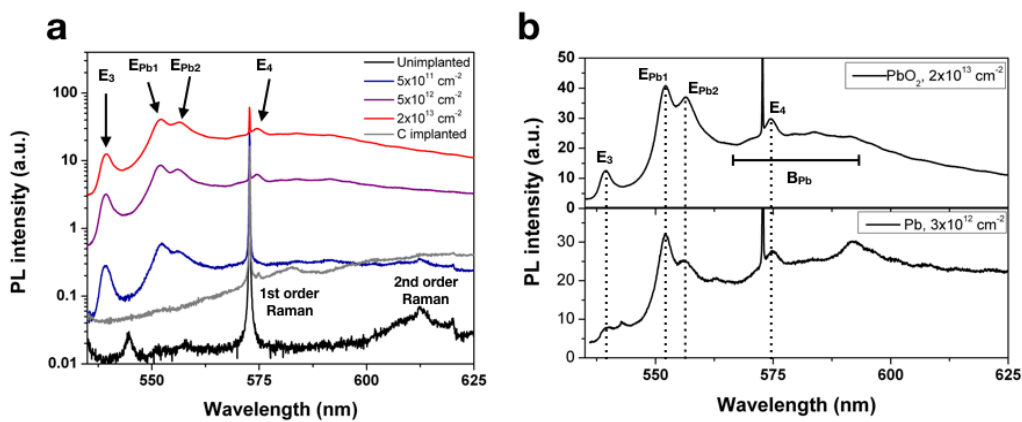

**Figure 2: a)** PL emission spectra of $PbO_2^-$-implanted diamond (Sample #1) at increasing ion fluences in the 5–200×$10^{11}$ cm$^{-2}$ cm$^{-2}$ range. Additional PL spectra acquired from different regions of the same substrate are reported for sake of comparison: unirradiated diamond (black line), and implanted with 15 keV C$^-$ (3×$10^{15}$ cm$^{-2}$ fluence, gray line) **b)** PL spectra of diamond implanted with 30 keV $PbO_2^-$ (Sample #1) 15 keV Pb$^-$ ions (Sample #2), respectively at 2×$10^{13}$ cm$^{-2}$ and 3×$10^{12}$ cm$^{-2}$ fluences.

A spectral feature similar to the $E_3$ line has previously been reported in natural diamond upon ion irradiation and high temperature (>1400 °C) annealing, and tentatively attributed to interstitial-carbon-related defect compkexes [16]. However, these previous observations were limited to the cathodoluminescence regime and the lack of clear evidence of its activity under PL excitation cannot rule out the attribution of the $E_3$ spectral line to a Pb-related defect. Only measurements at the single-photon emission level could provide an insight in the correlation of the $E_3$ PL peak with the $E_{Pb1}$-$E_{Pb2}$ emission lines.

Conversely, the $E_4$ emission is not reported in previous works, to the best of the authors' knowledge. Its spectral position within the $B_{Pb}$ band might suggest a similar role to the 640.8 nm peak observed for the GeV center [30]. According to this interpretation, the $E_4$ line would be related to the $E_{Pb1}$ emission by a phonon component of ~80 meV (i.e. ~705 cm$^{-1}$).

Attribution to Pb-related lattice defects. In order to attribute the spectral features reported in **Fig. 1** to Pb-related color centers, in Sample #1 we correlated the intensity of the PL emission with the implantation fluence in the 5–200×$10^{11}$ cm$^{-2}$ range. The corresponding PL spectra are shown in **Fig. 2a**. The afore-mentioned $E_{Pb1}$-$E_{Pb2}$ and $B_{Pb}$ features are absent in the spectrum acquired from a pristine reference region of the substrate, and conversely they exhibit a systematically increasing intensity at increasing implantation fluences. Furthermore, the absence of the the $E_{Pb1}$, $E_{Pb2}$ lines is apparent in a control spectrum acquired from a region of the same diamond substrate, implanted with 15 keV C$^-$ ions (3×$10^{15}$ cm$^{-2}$ fluence, **Fig. 2a**). It is also worth remarking that the $E_{Pb1}$ and $E_{Pb2}$ emission lines have never been observed in previous reports on diamond plates implanted with light (H, He) or carbon ions, as well as with other heavier ion species [5,6,16,36], thus indicating that these spectral features cannot be attributed to intrinsic radiation-induced defects and that the Pb impurities are directly involved in the formation of stable color centers.

Pb$^-$ vs $PbO_2^-$ ion implantation. The afore-mentioned data were acquired on a $PbO_2^-$-implanted substrate. We show in **Fig. 2b** that the reported spectral features ($E_{Pb1}$, $E_{Pb2}$, $E_3$, $E_4$) are left unchanged for both $PbO_2^-$ and Pb$^-$ ion implantations. This represents an additional experimental evidence that confirms their attribution to Pb-related optical centers, and that the presence of implanted O ions does not modify the relevant emission features. This result also has a significant technical relevance, as it confirms the possibility of fabricating Pb-related color centers with $PbO_2^-$ ions, i.e. by taking advantage of more stable ion beams and higher currents.

Temperature-dependent PL emission. In order to better investigate the fine structure of the spectral features observed in Pb-implanted diamond, PL spectra under 514 laser excitation were acquired at temperatures in the range comprised between 143 K (**Fig. 3a**) and 273 K (**Fig. 3b**). As typically observed in low-temperature PL experiments, the $E_{Pb1}$, $E_{Pb2}$, $E_3$ and $E_4$ PL peaks increase in intensity and reduce their spectral width with respect to room temperature. All the relevant Pb-related emission peaks displayed a blue shift at decreasing temperature (**Fig. 3c-3e**). This observation is in agreement with the attribution of the PL peaks to stable lattice defects, whose chemical bonds are shortened at decreasing temperature [37]. It is worth noting that the main emission peak $E_{Pb1}$ overlaps with the first-order Raman line (552.4 nm) under 514 nm excitation at 273 K, and its spectral position can be accurately determined only at temperatures below 233 K. As temperature decreases, its position reaches a value of 550.9 nm at 143 K; similarly, at the same temperature the $E_{Pb2}$ and the $E_4$ peak blue-shift to 554.9 nm and 573.7 nm **(Fig. 3c)**, respectively. None of these emissions exhibits a fine structure in the probed temperature range. Conversely, the $E_3$ peak reveals a splitting in two separate lines (537.5 nm and 538.2 nm at 143 K, **Fig. 3d**). An additional emission peak located at ~592 nm is superimposed to the second-order Raman scattering (~590-600 nm), and lies in the same spectral position of radiation-induced defects in diamond based on interstitial atoms [38]. Therefore an estimation of its intensity trend at decreasing temperatures and its unambiguous interpretation in relation to the $E_{Pb1}$-$E_{Pb2}$ lines was not possible.

Finally **Fig. 3a** shows the presence of two additional spectral lines (640.4 nm and 649.8 nm, $E_5$ and $E_6$ in the following) at 143 K, which could not be observed at room temperature. Consistently with the above reported peaks, these spectral features also exhibit a blue-shift at decreasing temperatures. Furthermore, they were also detected under 405 nm laser excitation, as will be discussed in the following.

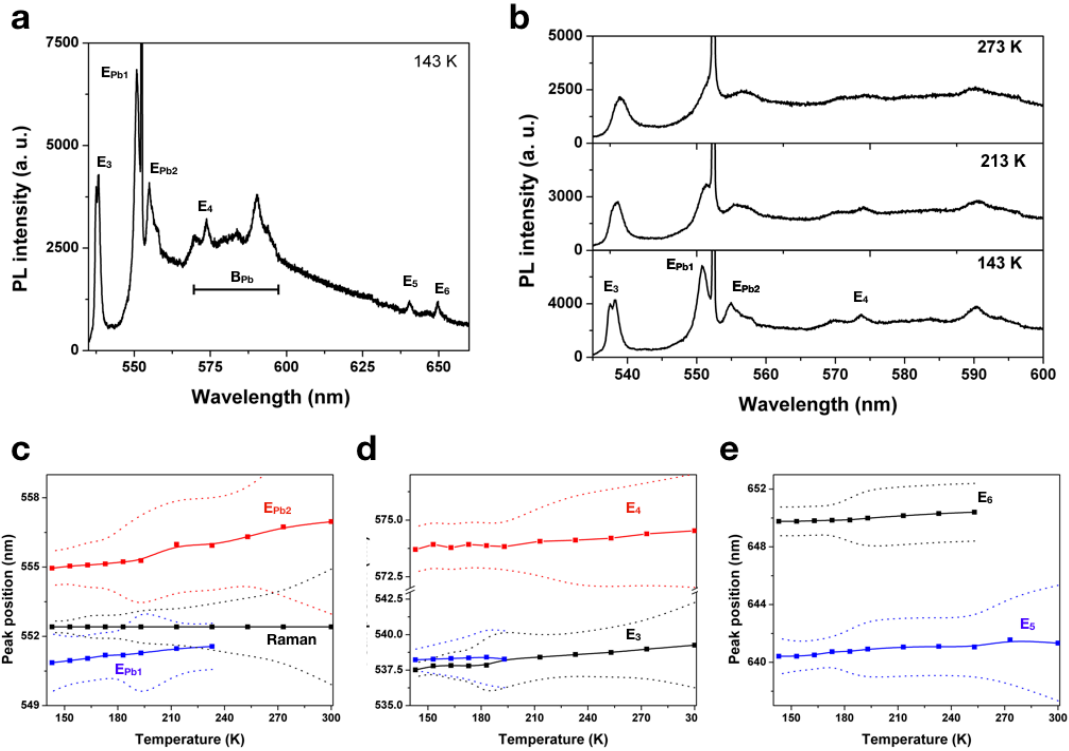

**Figure 3**: Temperature-dependent PL spectroscopy from PbO$_2^-$ implanted diamond: **a)** PL spectrum under 514 nm laser excitation at 143 K. **b)** Pb-related emission in the 535-600 nm spectral range at 273 K, 213 K, 143 K; temperature-dependent ZPL position of **c)** $E_{Pb1}$, $E_{Pb2}$, and first-order Raman emission; **d)** $E_3$ emission, with fine structure at T< 200 K, and $E_4$ emission; **e)** $E_5$, $E_6$ emissions. The dotted lines delimit the peaks' FWHM of the corresponding lines.

PL under 405 nm laser excitation. **Fig. 4** shows a typical PL spectrum in the 535-675 nm range acquired under 405 nm laser excitation. The $E_{Pb1}$, $E_{Pb2}$, $E_3$ peaks are still visible, although with a reduced intensity with respect to the $E_4$ spectral line. It is worth mentioning that $E_{Pb1}$ and $E_{Pb2}$ are convoluted due to the limited spectral resolution of the experimental system adopted for these measurements (~4 nm). The presence of the relevant Pb-related lines confirm the excitability of the optical centers at >3 eV excitation energy.

Furthermore, an intense emission from the previously mentioned $E_5$ (641.8 nm at room temperature) and $E_6$ (647.8 nm) peaks, together with an additional feature at 670.8 nm, is evident. These lines are not observable under 532 nm and 514 nm excitations at room temperature and barely detectable at 143 K under the same excitation wavelengths, while they instead represent the main spectral components of an articulated series of peaks in the 600-700 nm range under 405 nm excitation. While several radiation-induced defects have been studied in this spectral region [16], to the best of the authors' knowledge no simultaneous detection of both $E_5$ and $E_6$ has been reported in the scientific literature, to the best of the authors' knowledge. Therefore, their attribution to Pb-related defects cannot be ruled out, although such an attribution is still fairly tentative and undoubtedly requires further verification. The significant increase in the $E_5$ and $E_6$ lines intensity under higher excitation energy might also suggest their interpretation as a different charge state of the same defect(s) associated with the $E_{Pb1}$-$E_{Pb2}$ emission.

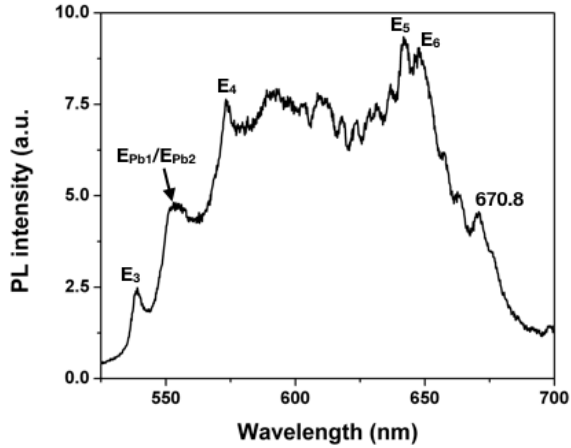

**Figure 4**: Ensemble PL emission from Sample #1 (PbO$_2^-$ implantation fluence: 2×10$^{13}$ cm$^{-2}$) acquired under 405 nm laser excitation.

The spectral separation of the E$_{Pb1}$ and E$_{Pb2}$ emission lines is apparent at room temperature. Should their attribution to the same optical center be confirmed by further studies at the single-photon emitter level, it is worth remarking that such energy difference (~16 meV) is smaller than the energy of quasi-local vibration expected for a Pb-related center (estimated as ~21 meV for $^{208}$Pb [16,32,39]) and therefore could be interpreted in terms of a significant splitting of the excited state levels of the defect [33].

Under this (yet to be demonstrated) hypothesis, the energy difference between the E$_{Pb1}$ and E$_{Pb2}$ spectral components at 143 K would correspond to a ~3900 GHz between the split levels of the excited state. Notably, this value lies in the same range as that identified for the energy splitting of the excited state of the SnV center (~3000 GHz [33]). Such similarity is somewhat surprising, as a significant increase in the ground-state splitting for the other group-IV-related color centers in diamond could be expected, the spin-orbit coupling constant increasing with the atomic number of the impurity [40].

An additional open question is represented by the attribution of the E$_3$ emission. On one hand, it cannot be ruled out that this peak corresponds to a secondary center, which anneals out at higher temperatures, similarly to the 593.5 nm emission in Sn-implanted diamond [32,33]. On the other hand, its splitting at low temperatures might also indicate a common origin from the very same Pb defect responsible for the E$_{Pb1}$ and E$_{Pb2}$ emissions. This hypothesis might be supported by the observation of a ~750 GHz splitting the E$_3$ emission, i.e. comparable with what observed for the ground state of the SnV center (~850 GHz, [33]). Conversely, according to this interpretation the significant energy difference (~55 meV) between the E$_{Pb1}$ and the E$_3$ peaks would imply the existence of additional physical effects, to be identified only by means of a dedicated modeling of the defect structure.

Conclusions. We reported on the formation of optically-active centers in diamond upon the introduction of Pb impurities by ion implantation. The emission intensity of the observed PL lines in the 535-675 nm spectral range strongly correlates with the implantation fluence, thus providing a solid indication that such emission originates from stable Pb-containing lattice defects. The number of emission peaks and the complexity of the PL spectra do not enable, on the basis of the current set of measurement, an unambiguous attribution of each one of the observed spectral features to specific defects and corresponding electronic configurations. The most intense emission peaks at room temperature are located at 552.1 nm (E$_{Pb1}$) and 556.8 nm (E$_{Pb2}$), and could possibly correspond to a significant splitting of the defect ground state, related to the high atomic number of the Pb impurity. The relation of such emission lines with additional spectral features, including a doublet at 539.4 nm and a peak at 574.5 nm, is currently unclear. Furthermore, two PL peaks at 641.8 and 647.8 nm (E$_5$, E$_6$), barely excitable under 514 nm excitation and conversely displaying an intense luminescence under 405 nm excitation, could consist of an alternative charge state of the very same defect responsible for the E$_{Pb1}$-E$_{Pb2}$ emission lines. On the basis of these results, a full understanding of the spectral features of Pb-related defects comparatively with the available reports on the other group-IV impurities seems not straightforward. A disambiguation on the photo-physical properties of these defects will require a thorough comparison with dedicated theoretical modeling and simulations of the defect's structure, additional spectral analysis on a wider temperature range and most importantly an unambiguous characterization of the centers at the single-photon emitter level.

These results represent a significant step towards completing the picture on the optical activity of diamond defects related to group IV impurities. Future studies on the defects' properties at the single-photon emitter level could lead to appealing perspectives in the fields of applied sciences and quantum information processing. Furthermore, from a fundamental point of view, the mapping and thorough understanding of a general pattern in the the opto-physical properties of color centers associated with impurities of the whole group IV could provide an important reference for the study of defects related to other chemical species.

## Acknowledgements

This research activity was supported by the following projects: "DIESIS" project funded by the Italian National Institute of Nuclear Physics (INFN) - CSN5 within the "Young research grant" scheme; Coordinated Research Project "F11020" of the International Atomic Energy Agency (IAEA); EMPIR Project. No. 17FUN06 "SIQUST" and 17FUN01 "BECOME" (the EMPIR initiative is co-funded by the EU H2020 and the EMPIR Participating States).

T.L., S.P. and J.M. gratefully acknowledge the support of VolkswagenStiftung. S.D. gratefully acknowledges the "Erasmus Traineeship 2016-2017" program for the financial support to the access to the ion implantation facilities of the University of Leipzig. P.O. acknowledges support from the project "Departments of Excellence" (L. 232/2016), funded by the Italian Ministry of Education, University and Research (MIUR).